\documentclass{article}
\usepackage{spconf,amsmath,graphicx,hyperref, amssymb}
\usepackage{cite} 
\usepackage{xcolor}
\usepackage{booktabs}
\usepackage{algorithm}
\usepackage{algorithmicx}
\usepackage{algpseudocode}
\usepackage{verbatim}
\usepackage{adjustbox}

\input{mysymbol.sty}

\title{Wireless Link Scheduling with Graph Signal Diffusion}
\name{Teresa Shang, Romina Garcia Camargo, Yi\u{g}it Berkay Uslu, Alejandro Ribeiro
}
\address{Dept. of Electrical and Systems Engineering \\ University of Pennsylvania \\ Philadelphia, PA, USA}
\begin{document}
\ninept
\maketitle
\begin{abstract}
We solve long-horizon wireless link scheduling with per-link minimum transmission constraints. Fair distribution of resources across links and time steps requires stochastic scheduling, motivating a diffusion framework. Via imitation learning, we train a diffusion model to sample scheduling decisions, conditioning generation on the network topology and constraint levels. The parameterization is achieved through a U-Graph Neural Network (U-GNN) to leverage the relationship between graphs and wireless networks for learning. Our approach achieves near-optimal performance, maximizing sum rate while distributing allocations across different links over time horizons. The U-GNN-based diffusion model generalizes to requirements and network sizes unseen during training, demonstrating the robustness of the approach.
\end{abstract}

\label{sec:intro}
\section{Introduction}
Resource allocation in wireless networks requires dynamic, channel-adaptive algorithms to guarantee user quality of service (QoS). A critical challenge in network management is wireless link scheduling: determining the active subset of links in a time-shared channel such that interference is minimal \cite{naderializadeh2014itlinq, shen2017fplinq, ti2016itlinq2,cui2018spatialdlforwls}. While the standard problem formulation maximizes instantaneous sum rate \cite{ti2016itlinq2, shen2017fplinq, cui2018spatialdlforwls, he2022gblinks, lee2021graphembeddingwls, zhao2023wlsusinggnns}, user experience aligns closer with long-term performance metrics, motivating optimization over extended time horizons \cite{camargo2026longhorizonwirelesslinkscheduling} (Section \ref{sec:longhorizon-wls}). 
Moreover, under per-link minimum transmission requirements, no fixed schedule can serve every link, making the problem inherently stochastic.

Instantaneous wireless link scheduling has been widely studied through optimization \cite{shen2017fplinq, naderializadeh2014itlinq, ti2016itlinq2, wu2013flashlinq} and learning-based techniques \cite{cui2018spatialdlforwls, he2022gblinks, lee2021graphembeddingwls, zhao2023wlsusinggnns, shan2024grlinq}. FPLinQ \cite{shen2017fplinq} is the state-of-the-art heuristic, based on iterative fractional programming to maximize sum rates. Learning-based approaches leverage the natural representation of wireless networks as graphs, often running imitation learning on FPLinQ \cite{he2022gblinks, zhao2023wlsusinggnns,lee2021graphembeddingwls}. 
Long-horizon wireless link scheduling is approached using state augmentation on Graph Neural Network (GNN) architectures \cite{calvo2024stateaugmented, uslu2026faststateaugmented, camargo2026longhorizonwirelesslinkscheduling}. Frameworks leverage Lagrangian dual dynamics to update the input of the GNN, obtaining stochastic scheduling decisions after an initial transient delay \cite{camargo2026longhorizonwirelesslinkscheduling}. To bypass these delays and directly sample high-quality allocations, we turn to generative diffusion models, which generate stochastic signals via a learned denoising process and have recently emerged as a powerful alternative for resource allocation tasks \cite{liang2025diffusion, xue2026joint, wu2025drloptimization, uslu2025generative}. Our contributions are as follows.


\begin{itemize}

\item[\textbf{(C1).}] We cast long-horizon link scheduling with per-link minimum transmission requirements as learning a stochastic scheduling policy. We instantiate this policy as a conditional diffusion model trained to imitate a dual gradient descent expert (Sections~\ref{sec:longhorizon-wls} and~\ref{sec:diffusion}).

\end{itemize}

In the dual gradient descent expert, the multipliers act as adaptive link weights that time-share the channel among links. Parameterizing the diffusion model with a U-Graph Neural Network (U-GNN) \cite{uslu2026generativediffusionmodelsstochastic}, a GNN-based encoder--decoder architecture that we adapt from power control to link scheduling, we generate scheduling decisions offline as graph signals over the communication network's conflict graph \cite{jain2003impact}. Generation is conditioned on current channel states at the network level and on each link's time-averaged minimum transmission requirement at the link level. We present a thorough evaluation of the framework in numerical simulations (Section \ref{sec:experiments}).


\begin{itemize}

\item[\textbf{(C2).}] The proposed diffusion-based policies generate stochastic scheduling decisions with near-optimal performance, outperforming existing methods (Section \ref{sec:exps-optimalandconstraints}).
\end{itemize}

 Out-of-distribution evaluations across varying minimum transmission requirements highlight the robustness of the learned policy. Moreover, GNN parameterizations inherit structural properties such as scalability and transferability \cite{uslu2026generativediffusionmodelsstochastic, schmidt2026stabilityflowmodelsgraph}, enabling transfer across network sizes.

\begin{itemize}

\item[\textbf{(C3).}] Trained on networks with $500$ links, diffusion-based link scheduling policies transfer their performance without retraining to networks with up to $5000$ links, while remaining faster than the expert at every evaluated size (Section \ref{sec:transferability}).

\end{itemize}

\label{sec:format}
\section{Long-horizon Wireless Link Scheduling}
\label{sec:longhorizon-wls}

Consider a wireless network with $N$ links that share a single channel over slotted time $t = 1, 2, \ldots$. Under the primary interference model~\cite{hajekb88, zhao2023wlsusinggnns}, two links conflict if they share a common device. We represent these conflicts by a \emph{conflict graph} whose nodes are the links and whose edges join conflicting links. Its adjacency matrix ${\bbA \in \{0,1\}^{N \times N}}$ therefore has $a_{ij} = 1$ if links $i$ and $j$ interfere with each other and $a_{ij} = 0$ otherwise, with $a_{ii} = 0$. 

Given $\bbA$ and a schedule $\bbx \in \{0,1\}^N$, a link $i$ transmits successfully if and only if it is scheduled while none of its conflicting links are scheduled, i.e., if $x_i = 1$ and $[\bbA \bbx]_i = 0$. Formally, we express the successful transmissions as
\begin{equation} \label{eq:rates}
    \bbr(\bbx, \bbA) = \bbx \odot [\mathbf{1}_N - \bbA \bbx]_+ \in \{0,1\}^N,
\end{equation}
\noindent where $\odot$ denotes the entrywise (Hadamard) product and $[\cdot]_+$ is the entrywise nonnegative projection. Note that $\bbr(\bbx, \bbA)$ always indicates an independent set of the conflict graph $\bbA$, since a link that transmits successfully has no scheduled neighbors.


Maximizing the successful transmissions $\mathbf{1}^\top_N \bbr(\bbx, \bbA)$ in a single slot is a maximum independent set problem. While repeating its solution across all slots maximizes the sum throughput, it serves none of the links outside the chosen set, as is the case for any stationary scheduling policy. To serve every link fairly, we impose per-link minimum transmission rate requirements $\bbdelta \in [0,1]^N$ on the long-term fraction of slots in which each link transmits successfully, and refer to the pair $(\bbA, \bbdelta)$ as the \emph{network state}.

Consider a \emph{stochastic policy} $\ccalD_\bbx(\cdot \mid \bbA, \bbdelta)$, which is a conditional distribution over schedules from which $\bbx_t$ is drawn independently at each slot $t$. Throughout, we omit the conditioning on $(\bbA, \bbdelta)$ when clear from context. An optimal policy $\ccalD_\bbx^\star$ maximizes the expected sum throughput in \eqref{eq:rates} while satisfying the minimum transmission requirements over all distributions on $\{0,1\}^N$:
\begin{align}\label{eq:problemformulation-expectation}
    \ccalD^\star_{\bbx} \in \argmax_{\ccalD_\bbx} \quad &\mathbb{E}_{\ccalD_\bbx
    }
    \big[ \mathbf{1}_N^\top \bbr(\bbx, \bbA)\big]\nonumber\\
    \text{subject to} \quad &\mathbb{E}_{\ccalD_\bbx
    }\big[\bbr(\bbx, \bbA)\big]\geq \bbdelta.    
\end{align}
\noindent 

Since at most one link in any clique of the conflict graph may transmit in a given slot, an optimal policy $\ccalD_\bbx^\star$ for \eqref{eq:problemformulation-expectation} samples independent sets. As such, these solution distributions are highly multi-modal, and solving for them over $2^N$ schedules is computationally prohibitive for varying large-scale network deployments.

We circumvent this by introducing a parameterization through $\bbtheta$ and learning a parameterized policy $\ccalD_\bbx(\bbtheta)$. This shifts our goal to finding the optimal parameterized conditional distribution $\ccalD_\bbx(\bbtheta^\star)$ that approximates the target stochastic policies,
\begin{equation}\label{eq:learningproblem}
    \bbtheta^\star \in \argmin_{\bbtheta}\,
    \mathbb{E}_{(\bbA,\bbdelta)\sim\ccalD_{\bbA,\bbdelta}}\big[
    D_{\mathrm{KL}}\big(\ccalD^\star_\bbx \,\|\, \ccalD_\bbx(\bbtheta)\big)\big],
\end{equation}
\noindent where $\ccalD_{\bbA,\boldsymbol{\delta}}$ denotes the training distribution of network states. Solving~\eqref{eq:learningproblem} requires only samples from $\ccalD^\star_\bbx$, and not its closed form. We generate the samples as trajectories of a dual descent expert and replace $\ccalD^\star_\bbx$ with their empirical distribution $\widehat{\ccalD}_\bbx$ (Section~\ref{sec:expert_alg}). 


We treat schedules as graph signals on the conflict graph and parameterize $\ccalD_\bbx(\bbtheta)$ as a diffusion model conditioned on the network state $(\bbA, \bbdelta)$, from which we sample scheduling decisions $\bbx \sim \ccalD_\bbx(\bbtheta^\star)$ (Section~\ref{sec:diffusion}).




\begin{figure*}[!t]
    \centering
    \includegraphics[width=\textwidth]{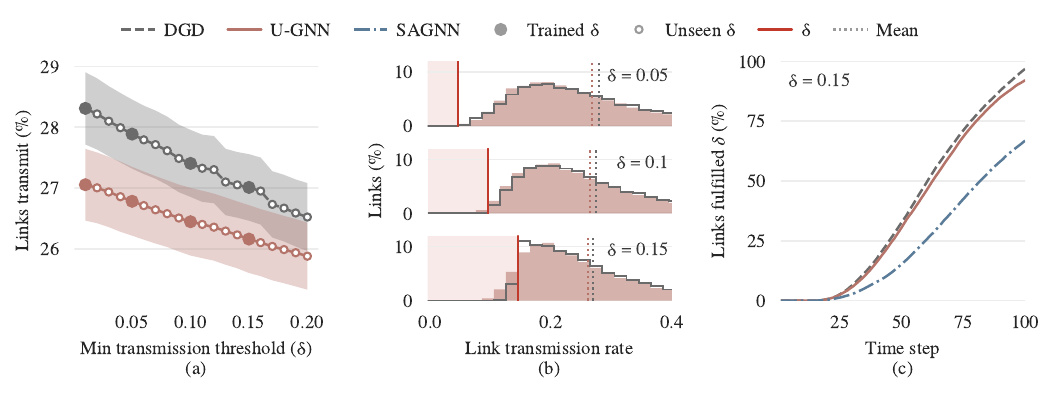}
    \par\vspace{-5pt}
    \caption{Performance and constraint satisfaction on 50 unseen networks for $T=100$ time steps. (a) Time-averaged sum rate as a function of $\delta$, showing mean and standard deviation across networks. Filled markers indicate thresholds used during training. (b) Per-link rate distributions of U-GNN-generated schedules (filled histograms) and expert samples (gray outlines). Solid red lines mark thresholds, shaded regions indicate violations, and dotted lines show means. (c) Percentage of links that have met their minimum transmission requirement by each time step, averaged across networks.}
    \label{fig:objective}
\end{figure*}

\subsection{Dual Gradient Descent}
\label{sec:expert_alg}

The expert solves~\eqref{eq:problemformulation-expectation} in the dual domain, where the search over distributions reduces to a sequence of problems over single schedules. Introducing nonnegative multipliers ${\boldsymbol{\lambda}\in\reals_+^N}$, with one multiplier for each link's minimum transmission constraint, the corresponding Lagrangian for~\eqref{eq:problemformulation-expectation} is
\begin{align}
    \ccalL(\ccalD_\bbx, \boldsymbol{\lambda}) =& \mathbf{1}_N^\top \mathbb{E}_{\ccalD_\bbx}\big[\bbr(\bbx, \bbA)\big] + \boldsymbol{\lambda}^\top(\mathbb{E}_{\ccalD_\bbx}\big[\bbr(\bbx, \bbA)\big] - \boldsymbol{\delta}).
    \label{eq:lagrangian_expectation}
\end{align}

We approximate the expectations in \eqref{eq:lagrangian_expectation} using empirical averages over a finite horizon of $T$ scheduling steps. The long-term Lagrangian can be reduced to an instantaneous formulation \cite{camargo2026longhorizonwirelesslinkscheduling}. Let $\boldsymbol{\lambda}_t \in \reals_+^N$ denote the vector of Lagrange multipliers at step $t \in [T]$. The corresponding \emph{instantaneous Lagrangian} is
\begin{align}
    \label{eq:instantaneous_lagrangian}
    \ccalL_{I}(\bbx, \boldsymbol{\lambda}_t)
    &= \mathbf{1}_N^\top \bbr(\bbx, \bbA) + \boldsymbol{\lambda}_t^\top\big(\bbr(\bbx, \bbA) - \bbdelta\big) \nonumber \\
    &= (\mathbf{1}_N + \boldsymbol{\lambda}_t)^\top \bbr(\bbx, \bbA) - \boldsymbol{\lambda}_t^\top \bbdelta.
\end{align}
At each scheduling step $t$, we select a schedule $\bbx_t$ that maximizes the instantaneous Lagrangian for the current multipliers $\boldsymbol{\lambda}_t$:
\begin{align}
\bbx_t \in \argmax_{\bbx} \ccalL_{I}(\bbx, \boldsymbol{\lambda}_t).
\end{align}
We then apply gradient descent to minimize the dual function, ${g(\boldsymbol{\lambda}_t) = \max_\bbx \ccalL_I(\bbx, \boldsymbol{\lambda}_t)}$, with respect to the multipliers:
\begin{align}
\boldsymbol{\lambda}_{t+1} = [\boldsymbol{\lambda}_t - \eta(\bbr(\bbx_t, \bbA) - \boldsymbol{\delta})]_+.
\label{eq:multiplier_update}
\end{align}
Here, $\eta > 0$ denotes the step size, and the rate residual ${\bbr(\bbx_t, \bbA) - \boldsymbol{\delta}}$ is a gradient of the dual function at $\boldsymbol{\lambda}_t$. 

Since $\boldsymbol{\lambda}_t^\top\bbdelta$ does not depend on $\bbx$, maximizing \eqref{eq:instantaneous_lagrangian} is equivalent to maximizing $(\mathbf{1}_N+\boldsymbol{\lambda}_t)^\top\bbr(\bbx, \bbA)$. Since $\bbr(\bbx, \bbA)$ always indicates an independent set, this maximization can be restricted without loss of optimality to the set $\ccalI \subseteq \{0,1\}^N$ of schedules whose active links form an independent set in the conflict graph, for which ${\bbr(\bbx, \bbA)=\bbx}$. Starting from $\boldsymbol{\lambda}_1=\bb0$, each step thus solves a maximum weighted independent set (MWIS) problem,
\begin{align}
    \bbx_t \in \argmax_{\bbx \in \ccalI}\; \bbw_t^\top \bbx,
    \qquad \bbw_t = \mathbf{1}_N + \boldsymbol{\lambda}_t.
    \label{eq:mwis}
\end{align}
We then update the multipliers using \eqref{eq:multiplier_update}, which determines the weights for the next step. Repeating this procedure for $T$ steps yields the expert sequences $\bbx_{1:T} \sim \widehat{\ccalD}_\bbx$. Since solving an MWIS problem at every step is costly (Figure~\ref{fig:size_transfer}(c)), we learn the empirical distribution with a diffusion model.

\section{Graph-Conditioned Diffusion Model}
\label{sec:diffusion}

A scheduling decision $\bbx$ assigns one value to each link and is thus a signal on the nodes of the conflict graph $\bbA$. We parameterize the learned policy $\ccalD_\bbx(\bbtheta)$ as a diffusion model of such graph signals~\cite{ho2020ddpm, uslu2026generativediffusionmodelsstochastic}, conditioned on the network state $(\bbA, \bbdelta)$. We distinguish the continuous diffusion variable from the binary schedule and let $\bbz_0 \in \reals^N$ denote an expert schedule for given $(\bbA, \bbdelta)$, i.e., $\bbz_0 \sim \widehat{\ccalD}_\bbx$. 

The forward diffusion process corrupts $\bbz_0$ with Gaussian noise over $M$ steps. The noisy signal at any step $m$ is given by
\begin{equation} \label{eq:noising}
    \bbz_m = \sqrt{\bar{\alpha}_m}\, \bbz_0 + \sqrt{1 - \bar{\alpha}_m}\, \boldsymbol{\epsilon},
    \qquad
    \boldsymbol{\epsilon} \sim \ccalN(\bb0, \bbI_N),
\end{equation}
where $\bar{\alpha}_m = \prod_{i=1}^m \alpha_i$ and $\alpha_m = 1 - \beta_m$ constitute the noise schedule chosen specifically so that $\bbz_{M} \approx \mathcal{N}(\bb0, \bbI)$ regardless of $\bbz_0$.

We learn a noise predictor $\boldsymbol{\epsilon}_{\bbtheta}(\bbz_m, m; \bbA, \bbdelta)$ that undoes this corruption by removing the noise $\boldsymbol{\epsilon}$ from $\bbz_m$, $m$ and network state $(\bbA, \bbdelta)$. We train it to minimize the DDPM objective~\cite{ho2020ddpm},
\begin{equation} \label{eq:ddpm}
    \bbtheta^\star \in \argmin_{\bbtheta} \
    \mathbb{E} \Big[ \big\| \boldsymbol{\epsilon}_{\bbtheta}(\bbz_m, m; \bbA, \bbdelta) - \boldsymbol{\epsilon} \big\|_2^2 \Big],
\end{equation}
where the expectation is over the training network states $(\bbA, \bbdelta)$ and noisy samples $\bbz_m$ formed via~\eqref{eq:noising} from independent draws of expert schedules $\bbz_0 \sim \widehat{\ccalD}_\bbx(\cdot \mid \bbA, \bbdelta)$, diffusion step $m \sim \mathrm{Unif}\{1, \ldots, M\}$, and noise $\boldsymbol{\epsilon} \sim \ccalN(\bb0, \bbI_N)$. We note that the objective in \eqref{eq:ddpm} is a variational bound on the negative log-likelihood of the expert samples~\cite{ho2020ddpm}, and thus a tractable surrogate for \eqref{eq:learningproblem}.

At inference, we sample from the trained model by drawing $\bbz_M \sim \ccalN(\bb0, \bbI_N)$ and running a deterministic DDIM sampler~\cite{song2021ddim}. Specifically, we estimate the clean sample at each step $m$ as
\begin{align}
    \hat{\bbz}_0 = \frac{\bbz_m - \sqrt{1 - \bar{\alpha}_m}\, \boldsymbol{\epsilon}_{\bbtheta^\star}(\bbz_m, m; \bbA, \bbdelta)}{\sqrt{\bar{\alpha}_m}},
    \label{eq:clean_estimate}
\end{align}
and use this estimate to obtain the next iterate as
\begin{align}
    \bbz_{m-1} = \sqrt{\bar{\alpha}_{m-1}}\, \hat{\bbz}_0 + \sqrt{1 - \bar{\alpha}_{m-1}}\, \boldsymbol{\epsilon}_{\bbtheta^\star}(\bbz_m, m; \bbA, \bbdelta), \label{eq:ddim_update}
\end{align}
\noindent for $m = M, \ldots, 1$ with $\bar{\alpha}_0 = 1$. A binary schedule $\bbx_t$ at each slot $t$ is obtained by thresholding the entries of the final iterate $\bbz_0$ at 1/2.

\subsection{U-Graph Neural Network Denoiser}
We parameterize $\boldsymbol{\epsilon}_{\bbtheta}$ with a U-Graph Neural Network (U-GNN) of~\cite{uslu2026generativediffusionmodelsstochastic}, a U-Net-style, permutation-equivariant backbone for graph signal diffusion that has been applied to power control over channel graphs~\cite{uslu2026generativediffusionmodelsstochastic, uslu2026graphsignaldiffusionmodels}. We use it without pooling as a symmetric encoder--decoder cascade of GNN blocks with long skip connections. Its core operation is the graph convolutional layer,
\begin{equation} \label{eq:gnn_layer}
    \bbY' = \bbPhi(\mathbf{Y}; \bbA, \ccalH) := \sigma \left( \sum_{k=0}^{K} \bbA^k \mathbf{Y} \bbH_k \right),
\end{equation}
\noindent which maps node features $\mathbf{Y} \in \reals^{N \times F}$ to $\mathbf{Y}' \in \reals^{N \times F'}$ by aggregating information within $K$ hops of the graph (shift operator) $\bbA$. Here, $\bbH_k \in \reals^{F \times F'}$ constitutes the learnable weights, and $\sigma$ is a nodewise nonlinearity, e.g., a ReLU. Cascading $L$ such layers yields a multi-layer GNN block, which we denote by the same symbol $\bbPhi(\cdot\,; \bbA, \ccalH)$, with $\ccalH$ now collecting the filter weights of all $L$ layers, for simplicity. 

The network stacks $B$ encoder blocks followed by $B$ decoder blocks, with filter weights $\ccalH^E_b$ and $\ccalH^D_b$ for the $b$th encoder and decoder block. From an initial (read-in) embedding of the input node features $\mathbf{Y}^E_0 = \Pi_{\mathrm{in}}(\bbz_m)$, they process the node features as
\begin{subequations} \label{eq:ugnn}
\begin{align}
    \mathbf{Y}^E_b &= \bbPhi \big( \mathbf{Y}^E_{b-1};\, \bbA,\, \ccalH^E_b \big), && b = 1, \ldots, B, \label{eq:ugnn_encoder} \\
    \mathbf{Y}^D_{b-1} &= \bbPhi \big( \big[ \mathbf{Y}^D_b,\, \mathbf{Y}^E_{b-1} \big];\, \bbA,\, \ccalH^D_b \big), && b = B, \ldots, 1, \label{eq:ugnn_decoder}
\end{align}
\end{subequations}
\noindent where $\mathbf{Y}^D_B = \mathbf{Y}^E_B$ connects the encoding and decoding paths, and $[\cdot, \cdot]$ denotes concatenation along the feature dimension. For readability, \eqref{eq:ugnn} omits the conditioning embeddings fused into each block input, which we describe below. A read-out layer then yields the noise prediction $\boldsymbol{\epsilon}_{\bbtheta}(\bbz_m, m; \bbA, \bbdelta) = \Pi_{\mathrm{out}}(\mathbf{Y}^D_0)$.

We particularize this architecture for our link scheduling problem as follows. The shift operator is the graph $\bbA$, while the conditioning consists of the per-link requirements $\bbdelta$. We embed $\bbdelta$ link-wise and the step $m$ by a sinusoidal encoding shared across links, and fuse both into the input of every block. A single model thus serves a range of requirement levels, including levels unseen in training (Section~\ref{sec:experiments}). 

\label{sec:pagestyle}
\section{Experiments} \label{sec:experiments}
We evaluate our diffusion model to determine its performance as a solution of \eqref{eq:problemformulation-expectation}.\footnote{Our code is publicly available at \url{https://github.com/tshang21/graph_diffusion_link_scheduling}.} In particular, we focus on the average sum rate achieved and the satisfaction of minimum transmission requirements. Our approach is compared against the Dual Gradient Descent (DGD) heuristic the model learns to imitate. We also compare the diffusion model against the State-Augmented GNN (SAGNN) approach in \cite{camargo2026longhorizonwirelesslinkscheduling}. We further characterize the generalization of our U-GNN-based model to unseen constraint levels and network sizes.

\begin{figure*}
    \centering
    \includegraphics[width=\textwidth]{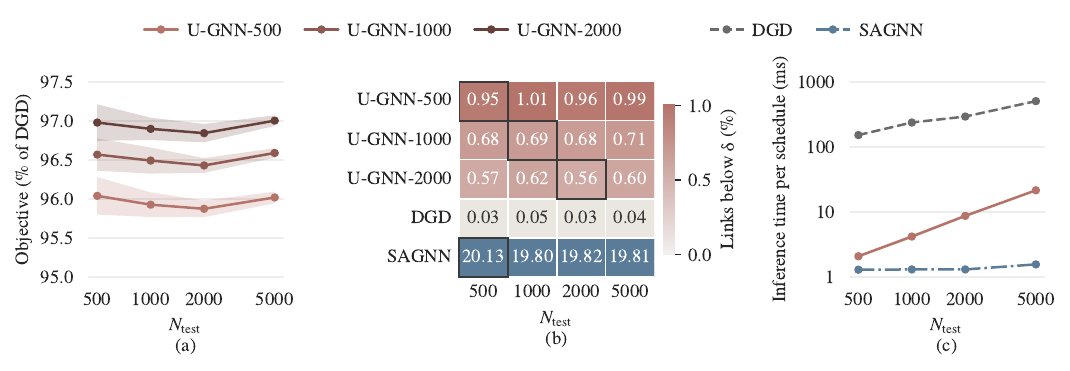}
    \par\vspace{-5pt}
    \caption{Transferability and scalability across test graph sizes $N_{\mathrm{test}}$. U-GNN-$N$ denotes a model trained on graphs containing $N$ links. (a) Objective value as a percentage of the DGD objective. (b) Percentage of links violating the minimum transmission requirement at $\delta=0.1$. Outlined cells indicate matching training and testing sizes. (c) Inference time per schedule averaged over 50 graphs.}
    \label{fig:size_transfer}
\end{figure*}

\subsection{Experimental Setup}
\label{sec:expsetup}
To model communication networks, we generate Random Geometric Graphs \cite{penrose2003random} by placing nodes on a regular grid, perturbing their positions with Gaussian noise to introduce irregularity, and connecting nodes within a fixed communication radius. We construct the corresponding conflict graphs, whose nodes represent links and edges encode interference constraints. These conflict graphs average ${N\simeq500}$ nodes, with a mean degree of 5.73 and a maximum degree of 14. Our main experiments use 100 training graphs and 50 testing graphs.

To accommodate the irregularity of conflict graphs, we use minimum transmission constraints $\bbdelta=\delta\mathbf{1}_N$. During training, we use ${\delta \in \{0.01,\allowbreak 0.05,\allowbreak 0.10,\allowbreak 0.15\}}$. These requirements are chosen to ensure the problem is feasible and the expert provides meaningful solutions. For each training graph and each $\delta$ value, we generate a sequence of scheduling decisions with the DGD algorithm (see Section~\ref{sec:expert_alg}) using MWIS solvers \cite{Lamm2019Eaxactly, 10.1007/s10732-017-9337-x}. 

We use the U-GNN architecture presented in \cite{uslu2026generativediffusionmodelsstochastic} with 512 channels and 4 encoder--decoder block pairs with skip connections. The model is trained for 1000 epochs with a learning rate of $10^{-4}$, for scale $N=500$ unless otherwise noted. For each generated graph, we randomly select 10 windows of 10 consecutive schedules from DGD rollouts with different requirements as training samples. This procedure allows our U-GNN to learn the temporal dynamics of scheduling decisions for multiple constraint requirements during training. At test time, the U-GNN generates 10 schedules at a time, for a total of $T=100$ steps.

\subsection{Objective and Generalization Across Constraints}
\label{sec:exps-optimalandconstraints}
As shown in Figure~\ref{fig:objective}(a), our U-GNN produces stochastic scheduling decisions that match DGD closely across evaluated $\delta$ values.
As $\delta$ increases, minimum transmission constraints become more influential, resulting in reduced sum rates. The shift in the distribution of schedules among links is observed in Figure \ref{fig:objective}(b). The lower tails shift towards higher transmission rates, reflecting the stricter requirements for links that transmit infrequently. 

As shown in Table~\ref{tab:min_violation}, the U-GNN keeps the average fraction of links violating their minimum transmission requirement below $1\%$ at all evaluated thresholds up to $\delta = 0.10$, including thresholds not seen during training. When requirements become stricter, the problem approaches infeasibility and constraint violation increases. 
These results demonstrate that U-GNN can sample schedules that combine high throughput with strong fairness performance across different constraint levels, without access to past schedules or signals from dual dynamics at inference. Figure~\ref{fig:objective}(c) further shows that the fraction of links meeting their requirement under U-GNN closely tracks that of DGD throughout the horizon. Generated schedules spread transmissions over time similarly to the expert, whereas SAGNN falls behind.




\begin{table}[t]
\centering
\caption{Minimum constraint violation (\%) across $\delta$.}
\label{tab:min_violation}
\setlength{\tabcolsep}{2.5pt}
\begin{adjustbox}{max width=\columnwidth}
\begin{tabular}{l cccccccccc}
\toprule
$\delta$ & 0.02 & 0.04 & 0.06 & 0.08 & 0.10 & 0.12 & 0.14 & 0.16 & 0.18 & 0.20 \\
\midrule
DGD & 0.00 & 0.00 & 0.00 & 0.01 & 0.03 & 0.31 & 1.43 & 5.59 & 14.84 & 24.76 \\
U-GNN & 0.11 & 0.07 & 0.32 & 0.71 & 0.95 & 4.49 & 6.41 & 16.89 & 20.96 & 31.33 \\
\bottomrule
\end{tabular}
\end{adjustbox}
\end{table}


\subsection{Transferability Across Network Sizes}
\label{sec:transferability}
Prior work has demonstrated the size transferability of graph-based architectures \cite{10508628, porras2026sizetransferability}, motivating our evaluation of U-GNN across network scales. We train separate models on conflict graphs with 500, 1000, and 2000 links, and evaluate each model on graphs with $N_{\text{test}}$ links at $\delta=0.1$. This setup assesses transfer to both smaller and larger graphs without retraining. 

As shown in Figure~\ref{fig:size_transfer}(a)--(b), each model maintains similar objective values and constraint violation rates across test sizes, including 5000-link graphs, which exceed all training sizes. Training on larger graphs consistently improves both metrics. These results demonstrate effective transfer across the evaluated graph sizes and suggest that training on larger graphs improves performance across network scales \cite{camargo2026scalabilitygraphneuralnetwork}.

Figure~\ref{fig:size_transfer}(c) shows that U-GNN requires less inference time per schedule than DGD at every evaluated graph size. SAGNN achieves the lowest inference times, but approximately $20\%$ of links violate their minimum transmission requirement at $\delta=0.1$, compared with approximately $1\%$ or less for U-GNN (Figure~\ref{fig:size_transfer}(b)). U-GNN thus offers a favorable balance among inference speed, objective value, and constraint satisfaction.

\section{Conclusions}

In this work, we address the wireless link scheduling problem with graph signal diffusion. Our results show that a learned scheduling distribution can capture the allocation of transmission opportunities induced by expert dual dynamics. Stochastic sampling from this distribution achieves a near-optimal long-horizon transmission objective with few minimum constraint violations. These findings support learning scheduling distributions as an effective approach to long-horizon resource allocation. 

An interesting future direction considers diverse minimum transmission requirements across links instead of uniform thresholds. Moreover, while the per-schedule inference time remains low, a smaller architecture may enable matching the runtime of Graph Neural Network approaches with performance trade-offs.

\label{sec:print}

\vfill\pagebreak

\bibliographystyle{IEEEbib}
\bibliography{strings,refs}

\end{document}